\documentclass[12]{article}
\usepackage{fleqn}
\usepackage{graphicx}
\begin{document}
\Large
\begin{center}{\bf
On the Pauli graphs of $N$-qudits
}
\end{center}
\vspace*{-.0cm}
\begin{center}
Michel Planat$^{\dag }$ and Metod Saniga$^{\ddag}$
\end{center}
\vspace*{-.3cm} \normalsize
\begin{center}
\vspace*{.1cm} $^{\dag}$Institut FEMTO-ST, CNRS, D\' epartement LPMO, 32 Avenue de
l'Observatoire\\ F-25044 Besan\c con, France\\
(michel.planat@femto-st.fr)

\vspace*{.1cm}
 and
\vspace*{.1cm}

$^{\ddag}$Astronomical Institute, Slovak Academy of Sciences\\
SK-05960 Tatransk\' a Lomnica, Slovak Republic\\
(msaniga@astro.sk)

\end{center}

\vspace*{-.1cm} \noindent \hrulefill

\vspace*{.1cm} \noindent {\bf Abstract}

\noindent A comprehensive graph theoretical and finite geometrical study of the commutation relations between the generalized Pauli operators of $N$-qudits is performed in which vertices/points correspond to the operators and edges/lines join commuting pairs of them. As per two-qubits, all basic properties and partitionings of the corresponding {\it Pauli graph} are embodied in the geometry of the generalized quadrangle of order two. Here, one identifies the operators with the points of the quadrangle and groups of maximally commuting subsets of the operators with the lines of the quadrangle.
The three basic partitionings are (a) a pencil of lines and a cube, (b) a Mermin's array and a bipartite-part and (c) a maximum independent set and the Petersen graph. These factorizations stem naturally from the existence of three distinct geometric hyperplanes of the quadrangle, namely a set of points collinear with a given point, a grid and an ovoid, which answer to three distinguished subsets of the Pauli graph, namely a set of six operators commuting with a given one,
a Mermin's square, and set of five mutually non-commuting operators, respectively.
The generalized Pauli graph for
multiple qubits is found to follow from symplectic polar spaces of order two, where maximal totally isotropic subspaces stand for maximal subsets of mutually commuting operators. The substructure of the (strongly regular) $N$-qubit Pauli graph is shown to be pseudo-geometric, i.\,e., isomorphic to a graph of a partial geometry. Finally, the (not strongly regular) Pauli graph of a two-qutrit system is introduced; here it turns out more convenient to deal with its dual in order to see all the parallels with the two-qubit case and its surmised relation with the generalized quadrangle $Q(4,3)$, the dual of $W(3)$. 
  \\ \\
{

\hspace*{1.37cm} Symplectic Polar Spaces -- Finite Projective (Ring) Geometries

\noindent
\hrulefill

\section{Introduction}
\noindent The intricate structure of commuting/non-commuting
relations between $N$-qubit observables may serve as a nice
illustration of the distinction between the quantum and the
classical and failure of classical ideas about measurements. A
deeper understanding of this structure is central to the
explanation of quantum peculiarities such as quantum
complementarity, quantum entanglement as well as other related
conceptual (or practical)  issues like no-cloning, quantum
teleportation, quantum cryptography and quantum computing, to
mention a few. Many ``strange features" of finite quantum
mechanics are linked with two important open theoretical
questions: finding complete sets of mutually unbiased bases
\cite{PlanatMUBs} and/or solving the Kochen-Specker theorem in
relevant dimensions \cite{Mermin}. Both problems are tricky and
difficult due to a large number of the observables involved.
Already for a two-qubit system, there are as many as fifteen
operators --- tensor products of the four Pauli matrices. This set
can be viewed as a graph if one regards the operators as vertices
and joins any pair of commuting ones by an edge. The two-qubit
{\it Pauli graph}, henceforth referred to as $\mathcal{P}[2,2]$,
is regular of degree six, that is, every observable commutes with
other six; one of its subgraphs, frequently termed as a Mermin's
square, has already been thoroughly studied due to its relevance
to a number of quantum ``paradoxes" \cite{Mermin,Planat1}. For
$N$-qubits ($N$-qutrits), $N > 2$, the corresponding graphs
$\mathcal{P}[2,N]$ ($\mathcal{P}[3,N]$) are endowed with $4^N-1$
($9^N-1$) vertices. One of their partitions features $2^N + 1$
($3^N + 1$) maximally commuting sets of $2^N-1$ ($3^N-1$)
operators each and is intimately related to the derivation of the
maximum sets of mutually unbiased bases in the corresponding dimensions
\cite{Lawrence1,Lawrence2}.

This paper aims at an
in-depth understanding of the properties of the $N$-qudit Pauli graphs by employing a number of novel graph theoretical and finite geometrical tools. It is organized as follows.
Sec.\,\ref{excerpts} first lists basic notions and definitions of graph theory and then introduces the relevant finite geometries. The latter start with the ubiquitous Fano plane, continue with other remarkable finite projective configurations (e.\,g., Pappus and Desargues) and related subspaces, and ends with more abstract and involved structures, such as generalized polygons and (symplectic) polar spaces. Sec.\,\ref{twoqubits} introduces the two-qubit Pauli graph and discusses its basic properties. The graph's three basic factorizations are then examined in very detail and their algebraic geometrical origin is pointed out: first, in terms of the three kinds of the geometric hyperplanes of the generalized quadrangle of order two, second in terms of the projective lines over the rings of order four and characteristic two residing in the projective line over $\mathcal{Z}_2^{2 \times 2}$ \cite{Saniga}.  
Sec.\,\ref{nqubits} discusses a self-similarity of the $N$-qubit graph; one shows that its structure is that of the symplectic polar spaces of order two \cite{Saniga4} and strongly regular graphs associated with them. Finally, Sec.\,\ref{twoqutrits} deals with some properties of the two-qutrit Pauli graph $\mathcal{P}[3,2]$ and muses about
possible finite geometry behind it.

\section{Graphs and geometry}
\label{excerpts}
\subsection{Excerpts from graph theory}
\label{intrograph}
\noindent
A graph $G$ consists of two sets, a non-empty set $V(G)$ of vertices and a set $E(G)$ of two element subsets of $V(G)$ called edges, the latter regarded as joins of two vertices. Alternatively, vertices are also called points and edges also lines \cite{Harary,Holton,Mohar}. Two distinct vertices of $G$ are called adjacent if there is an edge joining them; similarly, two distinct edges with a common vertex are called adjacent. If one vertex belongs to one edge both are said to be incident. The adjacency matrix $A=[a_{ij}]$ of a graph $G$ with $|V(G)| = v$ vertices is an $v \times v$ matrix in which $a_{ij}=1$ if the vertex $v_i$ is adjacent to the vertex $v_j$ and $a_{ij}=0$ otherwise. The degree $D$ of a vertex in a graph $G$ is the number of edges incident with it; a regular graph is a graph where each vertex has the same degree. A strongly regular graph is a regular graph in which any two adjacent vertices are both adjacent to a constant number of vertices, and any two non adjacent vertices are also both adjacent to a constant,
though usually different, number of vertices. The graph spectrum $spec(G)$ is composed of the eigenvalues (with properly counted multiplicities) of its adjacency matrix. For a regular graph, the largest eigenvalue equals the degree of the graph and the absolute value of any other eigenvalue is less than $D$.

A subgraph of $G$ is a graph having all of its vertices and edges in $G$. For any set $S$ of vertices of $G$, the induced subgraph, denoted  $\langle S \rangle $, is the maximal subgraph $G$ with the vertex set $S$.  A vertex and an edge are said to cover each other if they are incident. A set of vertices which cover all the edges of a graph $G$ is called a vertex cover of $G$, and the one with the smallest cardinality is called a minimum vertex cover. The latter induces a natural subgraph $G'$ of $G$ composed of the vertices of the minimum vertex cover and the edges joining them in the original graph. An independent set (or coclique) $I$ of a graph $G$ is a subset of vertices such that no two vertices represent an edge of $G$. Given the minimum vertex cover of $G$ and the induced subgraph $G'$, a maximum independent set $I$ is defined from all vertices not in $G'$. The set $G'$ together with $I$ partition the graph $G$.

Two graphs $G$ and $H$ are isomorphic (written $G \cong H$) if there exists a one-to-one correspondence between their vertex sets which preserves adjacency. An invariant of a graph $G$ is a number associated with $G$ which has the same value for any graph isomorphic to $G$. A complete set of invariants would determine a graph up to isomorphism, yet no such set is known for any graph. The most important invariants for a graph $G$ are the number of its vertices $v=|V(G)|$, the number of its edges $e=|E(G)|$, the degree at each vertex, its girth $g(G)$, i.\,e., the length of a shortest cycle (if any) in $G$, its diameter and its (vertex) chromatic number.  The distance between two points in $G$ is the length of the shortest path joining them, if any. In a connected graph, distance is a metric. A shortest path is called a geodesic and the diameter of a connected graph is the length of the longest geodesic. A coloring of a graph is an assignment of colors to its points so that no two adjacent points have the same color. A $c$-coloring of a graph $G$ uses $c$ colors. The chromatic number $\kappa (G)$ is defined as the minimum $c$ for which $G$ has a $c$-coloring.

Quite often the structure of a given graph can be expressed in a compact form, in terms of smaller graphs and operations on them. Graph union, graph product, graph composition and graph complement are a few \cite{Harary}. The complement $\widehat{G}$ of a graph $G$ has $V(G)$ as its vertex set, and two vertices are adjacent in $\widehat{G}$ if they are not in $G$. We will also need the concept of the line graph $L(G)$ of a graph $G$, i.\,e., the graph which has a vertex associated with each edge of $G$ and an edge if and only if the two edges of $G$ share a common vertex.

\subsection{Graphs and finite geometries}
\label{introgeometry}
\noindent
A finite geometry may be defined as a finite space $\mathcal{S}=\{P,L\}$ of points $P$ and lines $L$ such that certain conditions, or axioms, are satisfied \cite{Batten}. One of the simplest set of axioms are those defining the so-called Fano plane: (i) there are seven points and seven lines, (ii) each line has three points and (iii) each point is on three lines. The Fano plane is a member of several communities, some of them of great relevance to the structure of an $N$-qubit system. It is, first of all, a {\it near linear space}, that is a space such that any line has at least two points and two points are on at most one line. The Fano plane is also a {\it linear space} for which the second axiom \lq\lq at most" can be replaced by \lq\lq exactly". More generally, a {\it projective plane} is a linear space in which any two lines meet and there exists a set of four points no three of which lie on a line. The projective plane axioms are dual in the sense that they also hold by switching the role of points and lines. In a projective plane every point/line is incident with the same number $k+1$ of lines/points,
where $k$ is called the order of the plane. It has been long conjectured that a projective plane exists if and only $k$ is a power of a prime number and this conjecture was related to the existence of complete sets of mutually unbiased bases for $N$-qudits \cite{SPR}. The Fano plane is, in fact, the smallest projective plane, having order $k=2$. Projective planes of order $k$ can be constructed from 3-dimensional vector spaces over finite fields $\mathbf{F}_{k}$; such planes are necessarily Desarguesian, but there also exists non-Desarguesian planes which do not admit such a coordinatization.

The Fano plane belongs also to a large family of {\it projective configurations}, which consist of a finite set of points and a finite set of lines such that each point is incident with the same number of lines and each line is incident with the same number of points. Such a configuration may be denoted $(v_a, e_b)$, where $v$ stands for the number of points, $e$ for the number of lines, $a$ is the number of lines per point and $b$ the number of points per line. If the number of points equals the number of lines one simply denotes a configuration as $(v_a)$, although it is not, in general, unique. A configuration is said to be self-dual  if its axioms remain the same by interchanging the role of points and lines. The Fano plane is a configuration $(7_3)$. We will soon meet other two distinguished
projective configurations: the Pappus configuration $(9_3)$ and the Desargues configuration $(10_3)$. All the three configurations are self-dual. Any configuration may also be seen as a regular graph by regarding its points as  vertices and its lines as edges.

Recently, another class of finite geometries was found out to be of great relevance for two-qubits --- projective lines defined over finite rings instead of  fields \cite{Planat1,Saniga1,sploc1,sploc2}. Given an associative ring $R$ with unity and $GL(2,R)$, the general linear group of invertible two-by-two matrices with entries in $R$, a pair $(\alpha,\beta)$ is called admissible over $R$ if there exist $\gamma,\delta \in R$ such that $\left(
\begin{array}{cc}
\alpha & \beta \\
\gamma & \delta \\
\end{array}
\right) \in {\rm GL}_{2}(R)$. The projective line over $R$ is
defined as the set of equivalence classes of ordered pairs
$(\varrho \alpha, \varrho \beta)$, where $\varrho$ is a unit of
$R$ and $(\alpha, \beta)$ admissible \cite{bh1,Saniga2}. Such a
line carries two non-trivial, mutually complementary relations of
neighbor and distant. In particular, its two distinct points $X$:
$(\varrho \alpha, \varrho \beta)$ and $Y$: $(\varrho \gamma,
\varrho \delta)$ are called {\it neighbor} if $\left(
\begin{array}{cc}
\alpha & \beta \\
\gamma & \delta \\
\end{array}
\right) \notin {\rm GL}_{2}(R)$ and {\it distant} otherwise. The
corresponding graph takes the points as vertices and its edges
link any two mutually neighbor points. For $R=\mathbf{F}_k$, (the
graph of) the projective line lacks any edge, being an independent
set of cardinality $k+1$, or a $(k+1)$-coclique. Edges appear only
for a line over a ring featuring zero-divisors, and their number
is proportional to the number of zero-divisors and/or maximal
ideals of the ring concerned (see, e.\,g., \cite{Saniga1}--\cite{Saniga2}
for a comprehensive account of the structure of finite projective
ring lines). Projective lines of importance for our model will be,
as already mentioned in Sec.\,1, the line defined over the
(non-commutative) ring of full $2\times 2$ matrices with
coefficients in $\mathcal{Z}_2$, as well as the lines defined over
three distinct types of rings of order four and characteristic two
\cite{Saniga}.

A linear space such that any two-dimensional subspace of it is a projective plane is called a {\it projective space}. The smallest non trivial exemple (other than the Fano plane) is the binary three dimensional space $PG(3,2)$ of which two-dimensional subspaces are Fano planes.
A {\it generalized quadrangle} is a near linear space such that given a line $L$ and a point $P$ not on the line, there is exactly one line $K$ through $P$ that intersects $L$ (in some point $Q$) \cite{Payne}. A finite generalized quadrangle is said to be of order $(s,t)$ if every line contains $s+1$ points and every point is in exactly $t+1$ lines and
it is called {\it thick} if both $s>1$ and $t>1$; otherwise, it is called {\it slim}. If $s=t$, we simply speak of a quadrangle of order $s$. A generalized quadrangle of order $(s,1)$ or $(1,t)$ is called a grid or a dual grid, both being slim. The simplest thick  generalized quadrangle,
usually denoted as $W(2)$, is of order 2; it is a self-dual object featuring 15 points/lines and a cornerstone of our model.

Further concepts closely related to a projective space are those of a subspace and of a geometric hyperplane. A set of points  in a projective space is a subspace if and only if for any line $L$ the set contains no point, one point, or all the points of $L$. More restrictively, a geometric hyperplane $H$ of a finite geometry is a
set of points such that every line of the geometry either contains
exactly one point of  $H$, or is completely contained in $H$.

Last but not least, we need to introduce the concept of  a {\it polar space}. A polar space $S=\{P,L\}$ is a near-linear space such that for every point $P$ not on a line $L$, the number of points of $L$ joined to $P$ by a line equals either one (as for a generalized quadrangle) or to the total number of points of the line \cite{Batten}. A polar space of rank $N$ $(N \ge 2)$ can also be seen \cite{Tits} as a set $\{P\}$ of points, together with certain subsets, called subspaces, such that: (a) every subspace, together with its own subspaces, is isomorphic to the projective space $PG(d,q)$ over the finite field $\mathbf{F}_q$ and of dimension $d$ at most $N-1$,
(b) the intersection of two subspaces is a subspace, (c) for each point $P$ not in a subspace $R$ of dimension $N-1$, there is a unique subspace $S$ of dimension $N-1$ such that $R \cap S$ is $(N-2)$-dimensional, and (d) there are at least two disjoint subspaces of dimension $N-1$.
A polar space of rank two is a generalized quadrangle. A particular class of higher-rank, $N>2$, polar spaces called symplectic polar spaces are, as already outlined in \cite{Saniga4}, the geometries behind (strongly regular) multiple-qubit Pauli graphs (Sec.\,\ref{nqubits}).

\section{The Pauli graph of two-qubits}
\label{twoqubits}
\noindent
Let us consider the fifteen tensor products $\sigma_i \otimes \sigma_j$, $i,j \in \{1,2,3,4\}$ and $(i,j)\neq (1,1)$, of Pauli matrices $\sigma_i= (I_2,\sigma_x, \sigma_y,\sigma_z)$, where  $I_2=\left(\begin{array}{cc}1 & 0 \\0 & 1\\\end{array}\right)$, $\sigma_x=\left(\begin{array}{cc}0 & 1 \\1 & 0\\\end{array}\right)$, $\sigma_z=\left(\begin{array}{cc}1 & 0 \\0 & -1\\\end{array}\right)$ and $\sigma_y=i \sigma_x \sigma_z$, label them as follows $1=I_2 \otimes \sigma_x$, $2=I_2 \otimes \sigma_y$, $3=I_2 \otimes \sigma_z$, $a=\sigma_x \otimes I_2$, $4=\sigma_x \otimes \sigma_x$\ldots, $b=\sigma_y \otimes I_2$,\ldots , $c=\sigma_z \otimes I_2$,\ldots, and find the product and the commutation properties of any two of them --- as given in Table 1 and Table 2, respectively. Joining two distinct mutually commuting operators by an edge, one obtains the Pauli graph $\mathcal{P}[2,2]$ with incidence matrix as shown in Table 2. After removing the triple $\{a,b,c\}$ of the ``reference" observables, the incidence matrix can be cast into a remarkably compact form (Table 3) which makes use of three $3\times 3$ matrices: $O$ (the ``zero" matrix), $A$ (the identity matrix) and $\hat{A}$ (the matrix complementary to $A$).
The main invariants of $\mathcal{P}[2,2]$ and those of some of its most important subgraphs are listed in Table 4.
As it readily follows from Tables 1--3, $\mathcal{P}[2,2]$ is $6$-regular and, so, intricately connected with the complete graphs $K_n$, $n=5$, $6$ or $7$. First, one checks that $\mathcal{P}[2,2] \cong \hat{L}(K_6)$, i.\,e., it is isomorphic to the complement of line graph of $K_6$.  Next, computing its minimum vertex cover (Table 4), one recovers the Petersen graph $PG \equiv \hat{L}(K_5)$. Finally, $\mathcal{P}[2,2]$ is also found to be isomorphic to the minimum vertex cover of $\hat{L}(K_7)$. Now, we turn to remarkable partitionings/factorizations and the corresponding distinguished subgraphs of $\mathcal{P}[2,2]$.
\begin{table}[t]
\begin{center}
\footnotesize
\begin{tabular}{||l||rrr|r|rrr|r|rrr|r|rrr||}
\hline\hline
 $ $& $1$ & $2$ & $3$ & $a$ & $4$ & $5$ & $6$ & $b$ & $7$ & $8$ & $9$ & $c$ & $10$ & $11$ & $12$ \\
\hline
\hline
1& $0$ & $i3$ & $-i2$ & $4$ & $a$ & $i6$ & $-i5$ & $7$& $b$ & $i9$ & $-i8$ & $10$ & $c$ & $i12$ & $-i11$\\
2& $-i3$ & $0$ & $i1$ & $5$ & $-i6$ & $a$ & $i4$ & $8$& $-i9$ & $b$ & $i7$ & $11$ & $-i12$ & $c$ & $i10$\\
3& $i2$ & $-i1$ & $0$ & $6$ & $i5$ & $-i4$ & $a$ & $9$& $i8$ & $-i7$ & $b$ & $12$ & $i11$ & $-i10$ & $c$ \\
\hline
$a$& $4$ & $5$ & $6$ & $0$ & $1$ & $2$ & $3$ & $ic$& $i10$ & $i11$ & $i12$ & $-ib$ & $-i7$ & $-i8$ & $-i9$\\
\hline
4& $a$ & $i6$ & $-i5$ & $1$ & $0$ & $i3$ & $-i2$ & $i10$& $ic$ & $-12$ & $11$ & $-i7$ & $-ib$ & $9$ & $-8$\\
5& $-i6$ & $a$ & $i4$ & $2$ & $-i3$ & $0$ & $i1$ & $i11$& $12$ & $ic$ & $-10$ & $-i8$ & $-9$ & $-ib$ & $7$\\
6& $i5$ & $-i4$ & $a$ & $3$ & $i2$ & $-i1$ & $0$ & $i12$& $-11$ & $10$ & $ic$ & $-i9$ & $8$ & $-7$ & $-ib$ \\
\hline
$b$& $7$ & $8$ & $9$ & $-ic$ & $-i10$ & $-i11$ & $-i12$ & $0$& $1$ & $2$ & $3$ & $ia$ & $i4$ & $i5$ & $i6$\\
\hline
7& $b$ & $i9$ & $-i8$ & $-i10$ & $-ic$ & $12$ & $-11$ & $1$& $0$ & $i3$ & $-i2$ & $i4$ & $ia$ & $-6$ & $5$\\
8& $-i9$ & $b$ & $i7$ & $-i11$ & $-12$ & $-ic$ & $10$ & $2$& $-i3$ & $0$ & $i1$ & $i5$ & $6$ & $ia$ & $-4$\\
9& $i8$ & $-i7$ & $b$ & $-i12$ & $11$ & $-10$ & $-ic$ & $3$& $i2$ & $-i1$ & $0$ & $i6$ & $-5$ & $4$ & $ia$ \\
\hline
$c$& $10$ & $11$ & $12$ & $ib$ & $i7$ & $i8$ & $i9$ & $-ia$& $-i4$ & $-i5$ & $-i6$ & $0$ & $1$ & $2$ & $3$\\
\hline
10& $c$ & $i12$ & $-i11$ & $i7$ & $ib$ & $-9$ & $8$ & $-i4$& $-ia$ & $6$ & $-5$ & $1$ & $0$ & $i3$ & $-i2$\\
11& $-i12$ & $c$ & $i10$ & $i8$ & $9$ & $ib$ & $-7$ & $-i5$& $-6$ & $-ia$ & $4$ & $2$ & $-i3$ & $0$ & $i1$\\
12& $i11$ & $-i10$ & $c$ & $i9$ & $-8$ & $7$ & $ib$ & $-i6$& $5$ & $-4$ & $-ia$ & $3$ & $i2$ & $-i1$ & $0$\\
\hline\hline
\end{tabular}
\label{Algeb}
\caption{The product properties between any two Pauli operators of two-qubits; $0 \equiv I_2$.}
\end{center}
\end{table}
%
\begin{table}[h]
\begin{center}
\begin{tabular}{||l||rrr|r|rrr|r|rrr|r|rrr||}
\hline\hline
 $ $& $1$ & $2$ & $3$ & $a$ & $4$ & $5$ & $6$ & $b$ & $7$ & $8$ & $9$ & $c$ & $10$ & $11$ & $12$ \\
\hline
\hline
1& $0$ & $0$ & $0$ & $1$ & $1$ & $0$ & $0$ & $1$& $1$ & $0$ & $0$ & $1$ & $1$ & $0$ & $0$\\
2& $0$ & $0$ & $0$ & $1$ & $0$ & $1$ & $0$ & $1$& $0$ & $1$ & $0$ & $1$ & $0$ & $1$ & $0$\\
3& $0$ & $0$ & $0$ & $1$ & $0$ & $0$ & $1$ & $1$& $0$ & $0$ & $1$ & $1$ & $0$ & $0$ & $1$ \\
\hline
$a$& $1$ & $1$ & $1$ & $0$ & $1$ & $1$ & $1$ & $0$& $0$ & $0$ & $0$ & $0$ & $0$ & $0$ & $0$\\
\hline
4& $1$ & $0$ & $0$ & $1$ & $0$ & $0$ & $0$ & $0$& $0$ & $1$ & $1$ & $0$ & $0$ & $1$ & $1$\\
5& $0$ & $1$ & $0$ & $1$ & $0$ & $0$ & $0$ & $0$& $1$ & $0$ & $1$ & $0$ & $1$ & $0$ & $1$\\
6& $0$ & $0$ & $1$ & $1$ & $0$ & $0$ & $0$ & $0$& $1$ & $1$ & $0$ & $0$ & $1$ & $1$ & $0$ \\
\hline
$b$& $1$ & $1$ & $1$ & $0$ & $0$ & $0$ & $0$ & $0$& $1$ & $1$ & $1$ & $0$ & $0$ & $0$ & $0$\\
\hline
7& $1$ & $0$ & $0$ & $0$ & $0$ & $1$ & $1$ & $1$& $0$ & $0$ & $0$ & $0$ & $0$ & $1$ & $1$\\
8& $0$ & $1$ & $0$ & $0$ & $1$ & $0$ & $1$ & $1$& $0$ & $0$ & $0$ & $0$ & $1$ & $0$ & $1$\\
9& $0$ & $0$ & $1$ & $0$ & $1$ & $1$ & $0$ & $1$& $0$ & $0$ & $0$ & $0$ & $1$ & $1$ & $0$ \\
\hline
$c$& $1$ & $1$ & $1$ & $0$ & $0$ & $0$ & $0$ & $0$& $0$ & $0$ & $0$ & $0$ & $1$ & $1$ & $1$\\
\hline
10& $1$ & $0$ & $0$ & $0$ & $0$ & $1$ & $1$ & $0$& $0$ & $1$ & $1$ & $1$ & $0$ & $0$ & $0$\\
11& $0$ & $1$ & $0$ & $0$ & $1$ & $0$ & $1$ & $0$& $1$ & $0$ & $1$ & $1$ & $0$ & $0$ & $0$\\
12& $0$ & $0$ & $1$ & $0$ & $1$ & $1$ & $0$ & $0$& $1$ & $1$ & $0$ & $1$ & $0$ & $0$ & $0$\\
\hline\hline
\end{tabular}
\caption{The commutation relations between pairs of Pauli operators of two-qubits {\it aka} the incidence matrix of the Pauli graph $\mathcal{P}[2,2]$.
The symbol ``0"/``1" stands for non-commuting/commuting; although the diagonal should
feature 1's (every operator commutes with itself), we put there 0's for the reason which will become apparent from the text.}
\label{P22}
\end{center}
\end{table}
\begin{table}[h]
\begin{center}
\begin{tabular}{|r|r|r|r|}
\hline
$O$& $A$ & $A$ & $A$ \\
\hline
$A$& $O$ & $\hat{A}$ & $\hat{A}$ \\
\hline
$A$& $\hat{A}$ & $O$ & $\hat{A}$ \\
\hline
$A$& $\hat{A}$ & $\hat{A}$ & $O$ \\
\hline
\end{tabular}
\label{simpleP22}
\caption{Structure of the incidence matrix of $\mathcal{P}[2,2]$ after removal of the triple of operators $\{a,b,c\}$. }
\end{center}
\end{table}
\begin{table}[htb]
\begin{center}
\begin{tabular}{|r|r|r||r|r||r|r|}
\hline
 $G$&$\mathcal{P}[2,2]$&$PG \cong MVC$&$MS$&$BP$&$FP$&$CB$\\
 \hline
$v$ &$15$&$10$&$9$&$6$&$7$&$8$\\
 $e$&$45$&15&$18$&$9$&$9$&$12$\\
 $_{spec(G)}$&$_{\{-3^5,1^9,6\}}$&$_{\{-2^4,1^5,3\}}$&$_{\{-2^4,1^4,4\}}$&$_{\{-3,0^4,3\}}$&$_{\{-2,-1^3,1^2,3\}}$&$_{\{-3,-1^3,1^3,3\}}$\\
 $g(G)$&$3$&5&$3$&$4$&$3$&$3$\\
 $\kappa(G)$&$4$&3&$3$&$2$&$3$&$2$\\
 \hline
\end{tabular}
\end{center}
\label{invariantsP22}
\caption{The main invariants of the Pauli graph $\mathcal{P}[2,2]$ and its subgraphs, including its
minimum vertex covering $MVC$ isomorphic to the Petersen graph $PG$. For the remaining symbols, see the text.}
\end{table}
\begin{figure}[h]
\centerline{\includegraphics[width=9.0truecm,clip=]{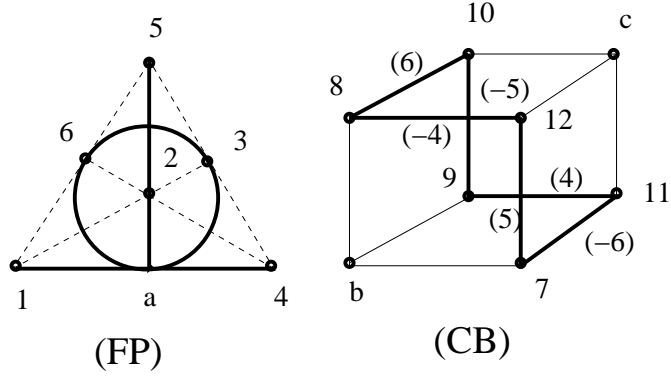}}
\caption{Partitioning of $\mathcal{P}[2,2]$ into a pencil of lines
in the Fano plane ($FP$) and a cube ($CB$). In $FP$ any two
observables on a line map to the third one on the same line. In
$CB$ two vertices joined by an edge map to points/vertices in
$FP$. The map is explicitly given for an entangled path by labels on the corresponding edges.}
\end{figure}
\begin{figure}[h]
\centerline{\includegraphics[width=9.0truecm,clip=]{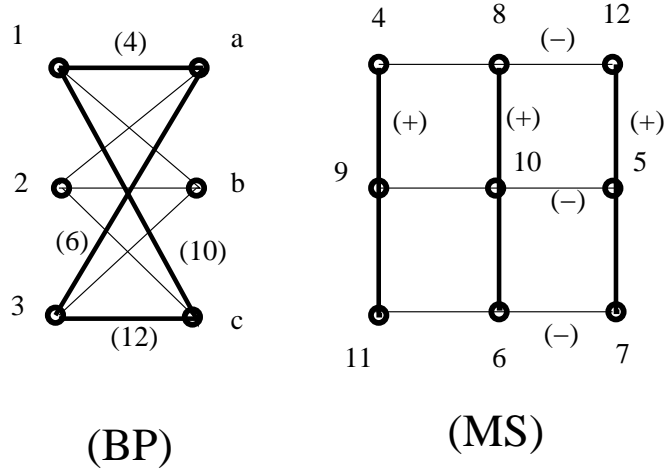}}
\caption{Partitioning of $\mathcal{P}[2,2]$ into an unentangled bipartite graph ($BP$) and a fully entangled Mermin square ($MS$). In $BP$ two vertices on any edge map to a point in $MS$ (see the labels of the edges on a selected closed path). In $MS$ any two vertices on a line map to the third one. Operators on all six lines carry a base of entangled states. The graph is polarized, i.e., the product of three observables in a row is $-I_4$, while in a column it is $+I_4$.}
\end{figure}
\begin{figure}[h]
\centerline{\includegraphics[width=6.6truecm,clip=]{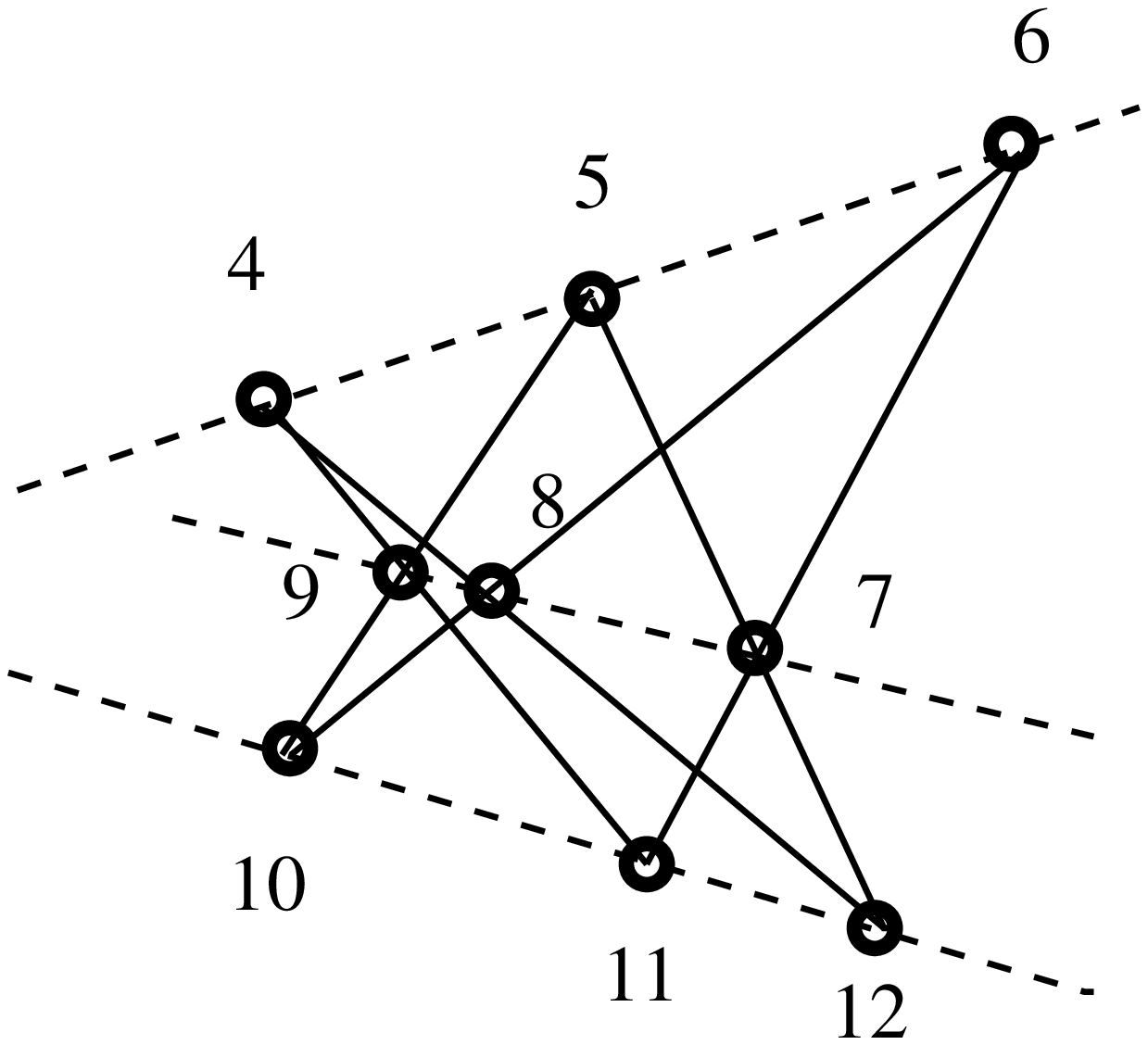}}
\caption{The Mermin square $MS$ viewed as a ``sub-Pappus" configuration; the Pappus configuration $(9_3)$ is obtained by adding the three extra lines (dotted).}
\end{figure}
\subsection{The ``Fano pencil" $FP$ and the cube $CB$}
\noindent
We shall first tackle the 7+8 partitioning of the graph which can, for example, be realized by the following subgraphs/subsets: $FP = \langle 1,2,3,a,4,5,6 \rangle$ and $CB = \langle b,7,8,9,c,10,11,12 \rangle$. The subgraph $FP$ can also be regarded as a line pencil in the Fano plane \cite{Planat1,Polster} as well as a hyperplane of $W(2)$ \cite{Saniga}; the number of choices for this partitioning is obviously equal to the number of the vertices of the full graph (see \cite{Planat1} for another choice). A $CB$ is also the generalized Petersen graph $G(4,1)$ \cite{Holton}. Employing Table 1, it is easy to observe that two vertices on one line of $FP$ map to the third one on the same line, i.e., $1.a=4$, $2.a=5$ and $3.a=6$. The three observables are found to share a common base of $4$-dimensional vectors;  for this particular choice, the lines in the Fano pencil $FP$ feature unentangled 2-qubit bases. In addition, an edge of $CB$ is mapped to a vertex of $FP$, e.\,g., $8.10=6$, $8.12=-4$, etc. In particular there is a closed path of length $6$ (shown with thick lines) in the cube graph $CB$ which features six bases of entangled states. It is worth mentioning here that  in \cite{Planat1} the projective lines over direct product of rings of the type $\mathcal{Z}_2^{\times n}$, $n=2,3,4$, were used to tackle this kind of partitioning. With these lines it was possible to grasp the structure of the two subsets, but not the coupling between them; to get a complete picture required employing a more abstract projective line with a more involved structure \cite{Saniga}.

\subsection{The Mermin square $MS$ and the bipartite part $BP$}
\noindent
We shall focus next on the 9+6 partitioning which can be illustrated, for example, by the subgraphs $BP=\langle 1,2,3,a,b,c\rangle$ and $MS=\langle 4,5,6,7,8,9,10,11,12\rangle$. The $BP$ part is easily recognized as the bipartite graph $K[3,3]$, while the $MS$ part is a $4$-regular graph.  There is a map from the edges of $BP$ to the vertices of $MS$, and a map from two vertices of a line in $MS$ to the third vertex on the same line. The bases defined by two commuting operators in $BP$ are unentangled. By contrast, operators on any row/column of $MS$ define an entangled base. A square/grid like the $MS$ was used by Mermin \cite{Mermin} --- and frequently referred to as a Mermin's square since then --- to provide a simple proof of the Kochen-Specker theorem in four dimensions. The proof goes as follows. One observes that the square is polarized in the sense that the product of three operators on any column equals $+ I_4$ (the $4 \times 4$ identity matrix), while the product of three observables on any row equals  $-I_4$. By multiplying all columns and rows one gets $-I_4$. This is, however, not the case for the eigenvalues of the observables; they all equal $\pm 1$ and their corresponding products always yield $+1$ because each of them appears in the product twice; once as the eigenvalue in a column and once as the eigenvalue in a row. The algebraic structure of mutually commuting operators thus contradicts that of their eigenvalues, which furnishes a proof of the Kochen-Specker theorem.
\begin{figure}[h]
\centerline{\includegraphics[width=9.2truecm,clip=]{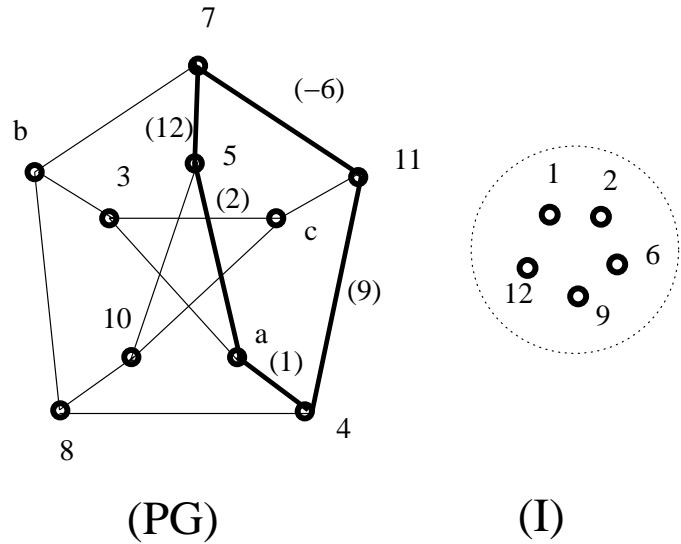}}
\caption{The partitioning of $\mathcal{P}[2,2]$ into a maximum independent set ($I$) and  the Petersen graph ($PG$), {\it aka} its minimum vertex cover. The two vertices on an edge of $PG$ correspond/map to a vertex in $I$ (as illustrated by the labels on the edges of a selected closed path).}
\end{figure}
\begin{figure}[h]
\centerline{\includegraphics[width=7.3truecm,clip=]{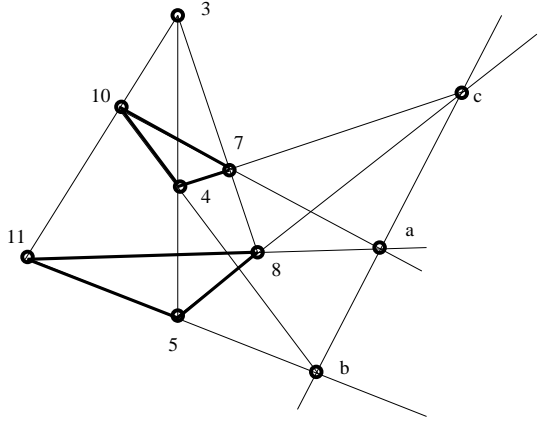}}
\caption{The complement of the Petersen graph viewed as the Desargues configuration; every line comprises three pairwise non-commuting operators  $o_k$, $o_l$, $o_m$, $k \neq l \neq m$, i.\,e., the operators obeying the rule $o_k.o_l=\pm i o_m$.}
\end{figure}
The $MS$ set is also recognized as a $(9_2,6_3)$ configuration for
any point is incident with two lines and any line is incident with
three points and does not change its shape if we reverse our
notation, i.\,e., join by an edge two mutually non-commuting
observables; in graph theoretical terms this means that the $MS$
equals its complement. It is also interesting to see that this
configuration sits inside the Pappus $(9_3)$ configuration (all
vertices and lines in Fig.\,3) by removing from the latter  the
three non-concurrent lines (the dotted ones). Last but not least,
it needs to be mentioned that the $MS$ configuration represents also
the structure of the projective line over the product ring
$\mathcal{Z}_2 \times \mathcal{Z}_2$ if we identify the points
sets of the two and regard edges as joins of mutually {\it distant}
points \cite{sploc1,sploc2}; it was precisely this fact that
motivated our in-depth study of projective ring lines
\cite{Saniga1,Saniga2} and finally led to the discovery of the
relevant geometries behind two- and multiple-qubit systems
\cite{Planat1,Saniga,Saniga4}.
\begin{figure}[h]
\centerline{\includegraphics[width=7.0truecm,clip=]{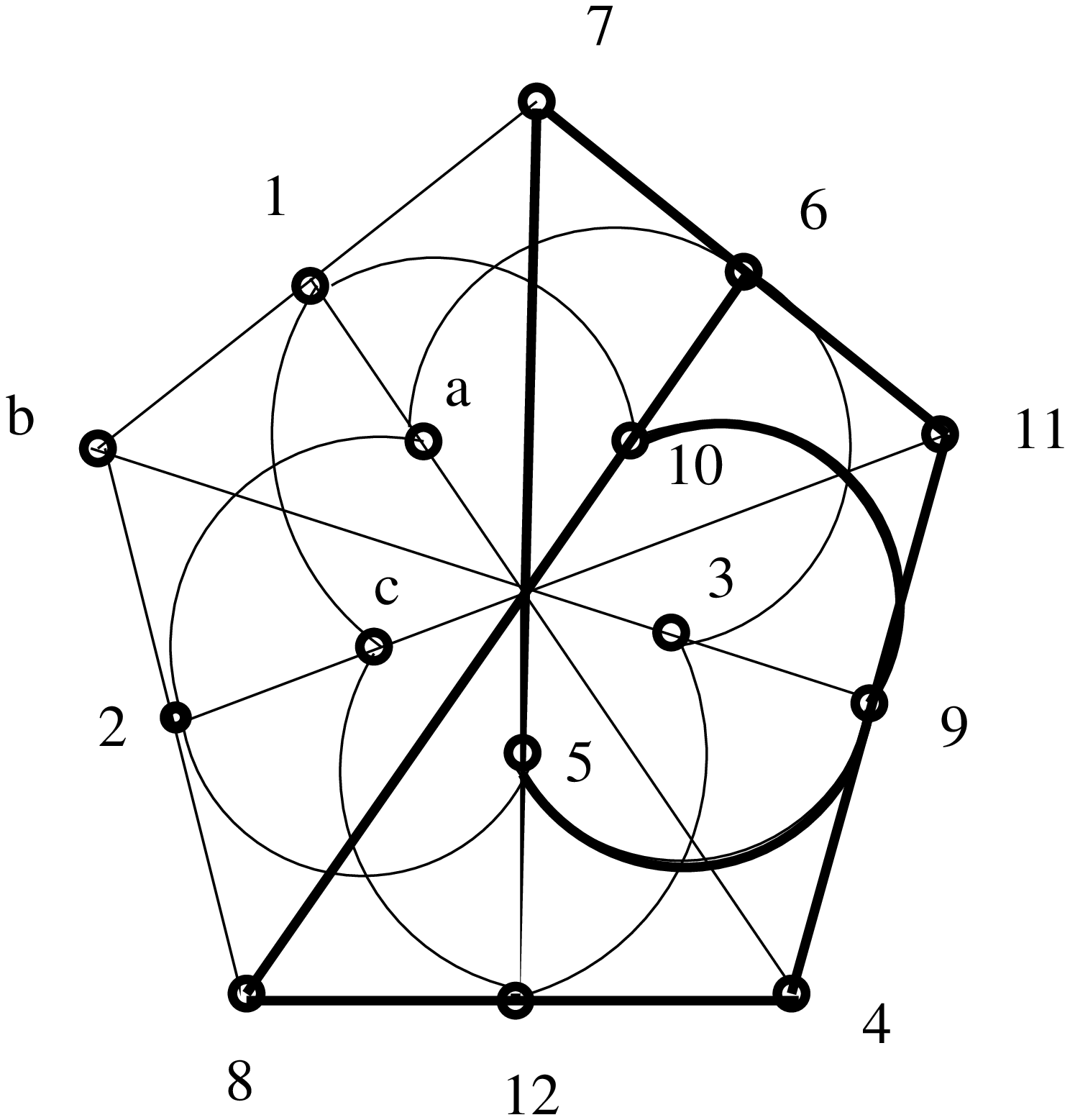}}
\caption{$W(2)$ as the {\it unique} underlying geometry of two-qubit systems. The Pauli operators correspond to the points and maximally commuting subsets of them to the lines of the quadrangle.  Three operators on each line have a common base; six out of fifteen such bases are entangled (the corresponding lines being indicated by boldfacing). }
\end{figure}
\subsection{The Petersen graph $PG$ and the maximum independent set $I$}
\noindent
The third fundamental partitioning of $\mathcal{P}[2,2]$ comprises a maximum independent set $I$ and the Petersen graph $PG$ \cite{Saniga}. This can be done in six different ways and one of them features $I=\langle 1,2,6,9,12\rangle$ and  $PG=\langle 3,a,4,5,b,7,8,c,10,11\rangle$.
As in the case of their cousins $CB$ and $BP$, the Petersen graph $PG$ admits a map of its edges into the vertices of the independent set $I$. Its complement, $\widehat{PG}$, can be viewed as a Desargues configuration $(10_3)$ (see Fig.\,5) whose points are the vertices of $PG$ and lines are triples of {\it non}-commuting observables $o_k$, $o_l$, $o_m$, $k \neq l \neq m$, $o_k.o_l=\pm i o_m$. The Desargues configuration is, like those of Fano and Pappus, self-dual.

\subsection{Finite projective algebraic geometry underlying $\mathcal{P}[2,2]$}
\subsubsection{$\mathcal{P}[2,2]$ as the generalized quadrangle of order two --- $W(2)$}
\noindent
At this point we have dissected $\mathcal{P}[2,2]$ to such an extent that we are ready to show the unique finite projective geometry hidden behind --- namely the {\it generalized quadrangle of order two}, $W(2)$ \cite{Saniga}. As already mentioned in Sec.\,2.2, $W(2)$ is the simplest thick generalized quadrangle endowed with fifteen points and the same number of lines, where every line features three points and, dually, every point is incident with three lines, and where every point is joined by a line (or, simply, collinear) with other six points \cite{Payne,Polster}. These properties can easily be grasped from the drawing of this object, dubbed for obvious reasons the doily, depicted in Fig.\,6; here, all the points are drawn as small circles, while lines are represented either by line segments (ten of them), or as segments of circles (the remaining five of them). To recognize in this picture $\mathcal{P}[2,2]$ one just needs to identify the fifteen points of $W(2)$ with our fifteen generalized Pauli operators as explicitly illustrated, with the understanding that {\it collinear} means {\it commuting} (and, so, {\it non}-collinear reads {\it non}-commuting); the fifteen lines of $W(2)$ thus stand for nothing but fifteen {\it maximum} subsets of three mutually commuting operators each. 

That $W(2)$ is indeed the right projective setting for $\mathcal{P}[2,2]$ stems also from the fact that it gives a nice geometric justification for all the three basic partitionings/factorizations of $\mathcal{P}[2,2]$. To see this, we just employ the fact that $W(2)$ features three distinct kinds of geometric hyperplanes \cite{Payne}: 1) a {\it perp}-set ($H_{cl}(X)$), i.\,e., a set of points collinear with a given point $X$, the point itself inclusive (there are 15 such hyperplanes);  2) a {\it grid} ($H_{gr}$) of nine points on six lines, {\it aka} a slim generalized quadrangle of order $(2,1)$ (there are 10 such hyperplanes); and 3) an {\it ovoid} ($H_{ov}$), i.\,e., a set of (five) points that has exactly one point in common with every line (there are six such hyperplanes). One then immediately sees \cite{Saniga} that a perp-set is identical with a Fano pencil, a grid answers to a Mermin square and, finally, an ovoid corresponds to a maximum independent set. Because of self-duality of $W(2)$, each of the above introduced hyperplanes has its dual, line-set counterpart. The most interesting of them is the dual of an ovoid, usually called a {\it spread},  i.\,e., a set of (five) pairwise disjoint lines that partition the point set; each of six different spreads of $W(2)$ represents such a pentad of mutually disjoint maximally commuting subsets of operators whose associated bases are {\it mutually unbiased} \cite{Planat1,Lawrence1}. It is also important to mention a {\it dual grid}, i.\,e., a slim generalized quadrangle of order $(1,2)$, having a property that the three operators on any of its nine lines share a base of {\it un}entagled states. It is straightforward to verify that these lines are defined by the edges of a $BP$; each of the remaining six lines (fully located in the corresponding/complementary $MS$) carries a base of {\it entangled} states (see Fig.\,6).

We shall finish this section with the following observation. A {\it triad} of a generalized quadrangle is an unordered triple of pairwise non-collinear points, with the common elements of the perp-sets of all the three points called its {\it centers} \cite{Payne}. $W(2)$ possesses two different kinds of triads: 1) those featuring {\it three} centers (e.\,g., the triple $\{b,5,11\}$), as well as 2) those which are {\it uni}centric (e.\,g., the triple $\{1,6,12\}$).  

\subsubsection{$\mathcal{P}[2,2]$ and the projective line over the full two-by-two matrix ring over $\mathcal{Z}_2$}
\label{ringline}
$W(2)$ is found as a {\it sub}geometry of many interesting projective configurations and spaces \cite{Payne,Polster}. We will now briefly examine a couple of such embeddings of $W(2)$ in order to reveal further intricacies of its structure and, so, to get further insights into the structure of the two-qubit Pauli graph.  

We shall first consider an embedding of $W(2)$ in the projective line defined over the ring $\mathcal{Z}_2^{2 \times 2}$ of full $2 \times 2$ matrices with $\mathcal{Z}_2$-valued coefficients,
\begin{equation}
 \mathcal{Z}_2^{2 \times 2}    \equiv \left\{ \left(
\begin{array}{cc}
\alpha & \beta \\
\gamma & \delta \\
\end{array}
\right) \mid ~ \alpha, \beta, \gamma, \delta \in \mathcal{Z}_2 \right\},
\end{equation}
because it was this projective ring geometrical setting where the relevance of
the structure $W(2)$ for two-qubits was discovered \cite{Saniga}.  
To facilitate our reasonings, we label the matrices of $\mathcal{Z}_2^{2 \times 2}$ in the following way
\begin{eqnarray}
&&~1' \equiv \left(
\begin{array}{cc}
1 & 0 \\
0 & 1 \\
\end{array}
\right),~2' \equiv \left(
\begin{array}{cc}
0 & 1 \\
1 & 0 \\
\end{array}
\right),~ 3' \equiv \left(
\begin{array}{cc}
1 & 1 \\
1 & 1 \\
\end{array}
\right),~
4' \equiv \left(
\begin{array}{cc}
0 & 0 \\
1 & 1 \\
\end{array}
\right), \nonumber \\
&&~5' \equiv \left(
\begin{array}{cc}
1 & 0 \\
1 & 0 \\
\end{array}
\right),~6' \equiv \left(
\begin{array}{cc}
0 & 1 \\
0 & 1 \\
\end{array}
\right),~ 7' \equiv \left(
\begin{array}{cc}
1 & 1 \\
0 & 0 \\
\end{array}
\right),~
8' \equiv \left(
\begin{array}{cc}
0 & 1 \\
0 & 0 \\
\end{array}
\right), \nonumber \\
&&~9' \equiv \left(
\begin{array}{cc}
1 & 1 \\
0 & 1 \\
\end{array}
\right),~10' \equiv \left(
\begin{array}{cc}
0 & 0 \\
1 & 0 \\
\end{array}
\right),~11' \equiv \left(
\begin{array}{cc}
1 & 0 \\
1 & 1 \\
\end{array}
\right),~12' \equiv \left(
\begin{array}{cc}
0 & 1 \\
1 & 1 \\
\end{array}
\right), \nonumber \\
&&13' \equiv \left(
\begin{array}{cc}
1 & 1 \\
1 & 0 \\
\end{array}
\right),~14' \equiv \left(
\begin{array}{cc}
0 & 0 \\
0 & 1 \\
\end{array}
\right),~15' \equiv \left(
\begin{array}{cc}
1 & 0 \\
0 & 0 \\
\end{array}
\right),~0' \equiv \left(
\begin{array}{cc}
0 & 0 \\
0 & 0 \\
\end{array}
\right),
\end{eqnarray}
and see that $\{1',2',9',11',12',13'\}$ are units (i.\,e., invertible matrices) and   $\{0',3',4',5',6',7',8',10',14',15'\}$ are zero-divisors (i.\,e., matrices with vanishing determinants), with 0' and 1' being, respectively, the additive and multiplicative identities of the ring.  Employing the definition of a projective ring line given in Sec.\,2.2, it
is a routine, though a bit cumbersome, task\footnote{See, for example,
\cite{Saniga1,sploc1} for more details about this methodology and  a number
of illustrative examples of a projective ring line.} to find out that the line over  $\mathcal{Z}_2^{2 \times 2}$ is endowed with 35 points whose coordinates, up to left-proportionality by a unit, read as follows
\begin{eqnarray}
&&(1',1'),~(1',2'),~(1',9'),~(1',11'),~(1',12'), (1',13'), \nonumber \\
&&(1',0'),~(1',3'),~(1',4'),~(1',5'),~(1',6'),~(1',7'),~(1',8'),~(1',10'),~(1',14'),~(1',15'), \nonumber \\
&&(0',1'),~(3',1'),~(4',1'),~(5',1'),~(6',1'),~(7',1'),~(8',1'),~(10',1'),~(14',1'),~(15',1'), \nonumber \\
&&(3',4'),~(3',10'),~(3',14'),~(5',4'),~(5',10'),~(5',14'),~(6',4'),~(6',10'),~(6',14').
\end{eqnarray}
Next, we pick up two mutually distant points of the line. Given the fact that
$GL(2,R)$ acts transitively on triples of pairwise distant points \cite{bh1}, the two points can, without any loss of generality, be taken to
be the points $U_{0}:=(1,0)$ and $V_{0}:=(0,1)$. The points of $W(2)$ are then those points of the line which are either simultaneously distant
or simultaneously neighbor to $U_{0}$ and $V_{0}$. The shared distant points are, in this particular representation, (all the) six points
whose both entries are units, 
\begin{eqnarray}
&&(1',1'),~(1',2'),~(1',9'), \nonumber \\
&&(1',11'),~(1',12'),~(1',13'),
\end{eqnarray}
whereas the common neighbors comprise (all the) nine points with both coordinates being zero-divisors,
\begin{eqnarray}
&&(3',4'),~(3',10'),~(3',14'), \nonumber \\
&&(5',4'),~(5',10'),~(5',14'),  \nonumber \\
&&(6',4'),~(6',10'),~(6',14'),
\end{eqnarray}
the two sets thus readily providing a ring geometrical explanation for a $BP+MS$ factorization of the algebra of the two-qubit Pauli operators, Fig.\,7, after the concept of mutually {\it neighbor} is made synonymous with that of mutually {\it commuting} \cite{Saniga}.
To see all the three factorizations within this setting it suffices to notice that the ring $\mathcal{Z}_2^{2 \times 2}$ contains as subrings all the {\it three} distinct kinds of rings of order four and characteristic two, viz. the (Galois) field $\mathbf{F}_4$, the local ring 
$\mathcal{Z}_2[x]/\langle x^2 \rangle$, and the direct product ring $\mathcal{Z}_2 \times \mathcal{Z}_2$ \cite{mcd}, and check that the corresponding lines can be identified with the three kinds of geometric hyperplanes of $W(2)$ as shown in Table\,5 \cite{Saniga}.
\begin{figure}[t]
\centerline{\includegraphics[width=11.3truecm,clip=]{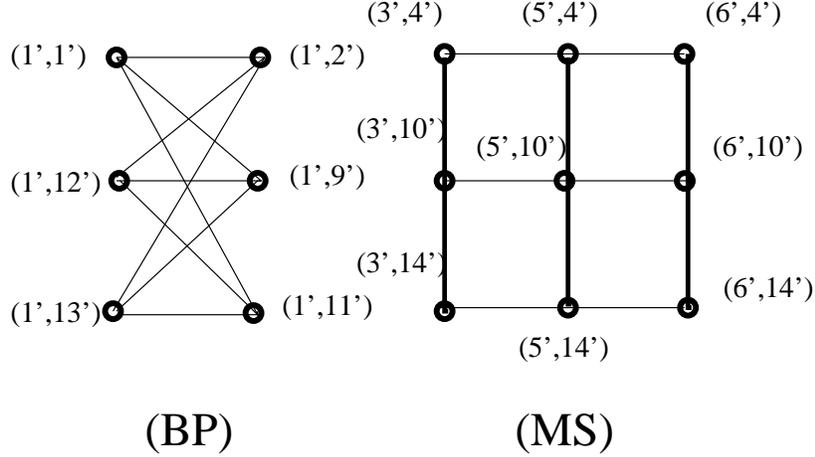}}
\caption{A $BP+MS$ factorization of $\mathcal{P}[2,2])$ in terms of the points of the subconfiguration of
the projective line over the full matrix ring $\mathcal{Z}_2^{2 \times 2}$; the points of the $BP$ have both coordinates units, whilst those of the $MS$ feature in both entries zero-divisors. The ``polarization" of the Mermin square is in this particular ring geometrical setting expressed by the fact that each column/row is characterized by the fixed value of the the first/second coordinate. Compare with Fig.\,2.}
\end{figure}
\begin{table}[hb]
\begin{center}
\caption{Three kinds of the distinguished subsets of the
generalized Pauli operators of two-qubits ($\mathcal{P}[2,2])$) viewed either as the geometric
hyperplanes in the generalized quadrangle of order two ($W(2)$) or as
the projective lines over the rings of order four and
characteristic two residing in  the projective line over $\mathcal{Z}_2^{2 \times 2}$.}
\vspace*{0.4cm}
\begin{tabular}{llll}
\hline \hline
\vspace*{-.3cm} \\
$\mathcal{P}[2,2]$ & set of five mutually    & set of six operators  & nine operators of a \\
   & non-commuting operators  & commuting with a given one & Mermin's square\\
   $W(2)$ & ovoid  &   perp-set$\setminus$\{reference point\} &   grid \\
Proj. Lines over & $\mathbf{F}_4 \cong \mathcal{Z}_2[x]/\langle x^2 + x + 1 \rangle$  &      $\mathcal{Z}_2[x]/\langle x^2 \rangle$  &   $\mathcal{Z}_2 \times \mathcal{Z}_2 \cong \mathcal{Z}_2[x]/\langle x(x+1) \rangle$  \\
\vspace*{-.3cm} \\
\hline \hline
\end{tabular}
\end{center}
\end{table}
\begin{figure}[ht]
\centerline{\includegraphics[width=8.6truecm,clip=]{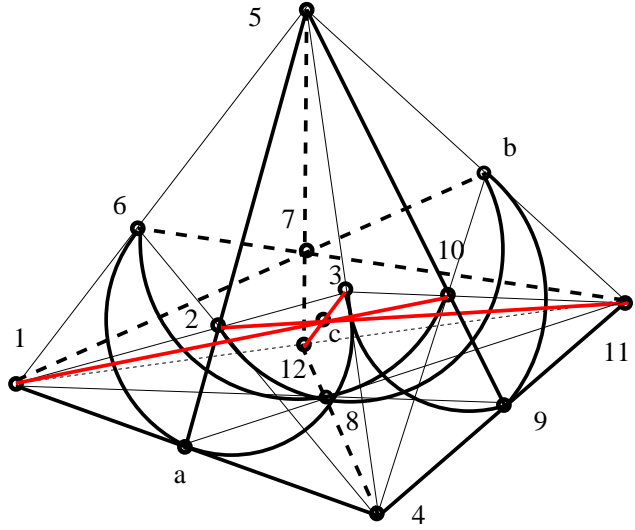}}
\caption{An illustration of an embedding of the generalized quadrangle $W(2)$ (and thus of the associated Pauli graph $\mathcal{P}[2,2])$ into the projective space $PG(3,2)$. The points of $PG(3,2)$ are the four vertices of the tetrahedron, its center, the four centers of its faces and the six centers of its edges; the lines are the six edges of the tetrahedron, the twelve medians of its faces, the four circles inscribed in the faces, the three segements linking opposite edges of the tetrahedron, the four medians of the terahedron and, finally, six circles located inside the tetrahedron \cite{Polster}. The fifteen points of $PG(3,2)$ correspond to the fifteen Pauli operators/vertices of $\mathcal{P}[2,2]$. All the thirty-five lines of the space carry each a triple of operators
$o_k$, $o_l$, $o_m$, $k \neq l \neq m$, obeying the rule $o_k.o_l = \mu o_m$; the operators located on the fifteen totally isotropic lines belonging to $W(2)$ yield $\mu=\pm 1$, whereas those carried by the remaining twenty lines (not all of them shown) give $\mu=\pm i$.}
\end{figure}

The other embedding of $W(2)$ to be briefly dealt with is the one into the projective space, $PG(3,2)$, as illustrated in Fig.\,8. This embedding is, in fact, a very close ally of the previous one due to a remarkable bijective correspondence between the points of the line over $\mathcal{Z}_2^{2 \times 2}$ and the lines of $PG(3,2)$ \cite{Thas}. $W(2)$ and $PG(3,2)$ are identical as the point sets, 
whilst the fifteen lines of $W(2)$ are so-called totally isotropic lines with respect to a symplectic polarity of $PG(3,2)$ (Sec.\,4.2). 

\section{The Pauli graph of $N$-qubits}
\label{nqubits}
\noindent
Following the same strategy as in the preceding section, we find out that
the $4^3-1=63$ tensor products $\sigma_i \otimes \sigma_j \otimes \sigma_k$, $[i,j,k=1,2,3,4$, $(i,j,k)\neq (1,1,1)]$ form the vertices and their commuting pairs the edges of a regular graph of degree $30$, $\mathcal{P}[2,3]$, with spectrum $\{-5^{27},3^{35},30\}$. The corresponding incidence matrix can also be cast into a compact tripartite form, Table 6, after the reference points $a_3=\sigma_x \otimes I_2 \otimes I_2$, $b_3=\sigma_y \otimes I_2 \otimes I_2$ and $c_3=\sigma_z \otimes I_2 \otimes I_2$ have been omitted. This matrix looks very much the same as its two-qubit counterpart (Table 3), save for the fact that  now all the submatrices are of rank $15 \times 15$. As in the two-qubit case, the matrix $A_3$ can simply be viewed as the join of $O_3$ and the unit matrix $I_8$. The same self-similarity pattern interrelating the incidence matrices of $(N+1)$- and $N$-qubit systems is found for any $N$.

As for the two-qubit incidence matrix, one of the most natural factorizations of the three-qubit matrix consists of the first block $O_3$ and a larger square block $\mathcal{M}_3$, of cardinality $45$,  containing $O_3$ and $\hat{A}_3$. The latter block is self-complementary, as is its two-qubit counterpart, a Mermin square; it represents a regular graph of degree $22$ and spectrum $\{-5^{10},-3^9,-2^2,1^5,3^{18},22\}$. The structure of this block is very intricate: it can be recovered again by removing from the reduced incidence matrix shown in Table 6 the first triple of points and all the reference points (of the type $a$, $b$ and $c$, see Table 2) of the parent scale, i.\,e., an extra set of $3+3 \times 4=15$ \lq\lq pseudo-reference" points of the ``daughter" scale.
\begin{table}[h]
\begin{center}
\begin{tabular}{|r|r|r|r|}
\hline
$O_3$& $A_3$ & $A_3$ & $A_3$ \\
\hline
$A_3$& $O_3$ & $\hat{A}_3$ & $\hat{A}_3$ \\
\hline
$A_3$& $\hat{A}_3$ & $O_3$ & $\hat{A}_3$ \\
\hline
$A_3$& $\hat{A}_3$ & $\hat{A}_3$ & $O_3$ \\
\hline
\end{tabular}
\label{simpleP22}
\caption{The incidence matrix of $\mathcal{P}[2,3]$ after removal of the triple of reference points (compare with Table\,3).}
\end{center}
\end{table}

After a closer look at $\mathcal{M}_3$, one reveals in it three subsets isomorphic to the Mermin square of two-qubits (Fig.\,2 and/or Fig.\,7), from which we can form doubles ($18$ points) and triples (27 points) having spectra $\{-3^4,-1^8,0,3^4,8\}$ and $\{-3^{12},0^6,3^8,12\}$, respectively; the graph of the latter bears number $105$ in the list of graphs with few eigenvalues given in \cite{Evandam}. 
One can also form $m$-tuples of the ``generalized" Mermin square of size $m =1, 2,3,4$ using the ``entangled" subset $\mathcal{E}$ located in the first block $O_3$ and the extra $MS$ copies from $\mathcal{M}_3$, to get another interesting blocks $\mathcal{E}\cup MS$,   $\mathcal{E}\cup (2 \times MS)$ and $\mathcal{E}\cup (3 \times MS)$ and the associated graphs with  spectra  $\{-3^4,-1^9,3^4,9\}$,
$\{-3^{12},0^5,3^8,3(2\pm \sqrt{6})\}$ and $\{-5^{4},-3^{12},0^2,1^4,3^{12},8\pm \sqrt{91}\}$, respectively.

\subsection{Rank $N$ symplectic polar spaces behind the $N$-qubit Pauli graphs}
\noindent
The geometry underlying higher order qubits \cite{Saniga4} can readily be hinted from the observation that our doily $W(2)$, embodying the two-qubit operators' algebra, is the lowest rank representative of a big family of {\it symplectic} polar spaces of order two.

A symplectic polar space (see, e.\,g.,  \cite{Tits,Cameron,Ball} for more
details) is a $d$-dimensional vector space
over a finite field $\mathbf{F}_q$, $V(d, q)$, carrying a non-degenerate bilinear
alternating form. Such a polar space,
usually denoted as $W_{d -1}(q)$,  exists only if $d=2N$, with $N$ being its
rank. A subspace of $V(d, q)$ is called totally isotropic if
the form vanishes identically on it. $W_{2N-1}(q)$ can then be
regarded as the space of totally isotropic subspaces of $PG(2N-1,
q)$ with respect to a symplectic form, with its maximal totally isotropic subspaces, called also generators $G$, having dimension $N - 1$. For $q=2$, this
polar space contains
\begin{equation}
|W_{2N-1}(2)| = | PG(2N-1, 2)| = 2^{2N} - 1 = 4^{N} - 1
\end{equation}
points and
$ (2+1)(2^{2}+1) \ldots (2^{N}+1)$
generators. A spread $S$ of $W_{2N-1}(q)$ is a set of generators partitioning
its points.  The cardinalities of a spread and a generator of
$W_{2N-1}(2)$ read
\begin{equation}
|S| = 2^{N} + 1
\end{equation}
and
\begin{equation}
|G| = 2^{N} - 1,
\end{equation}
respectively. Finally, it needs to be mentioned that two
distinct points of $W_{2N-1}(q)$ are called perpendicular if they
are joined by a line; for $q=2$, there exist
\begin{equation}
\#_{\Delta} = 2^{2N-1}
\end{equation}
points that are {\it not} perpendicular to a given point.

Now, in light of Eq.\,(6), we can identify the Pauli operators of $N$-qubits
with the points of $W_{2N-1}(2)$. If, further, we identify the
operational concept ``commuting" with the geometrical one
``perpendicular," from Eqs.\,(7) and (8) we readily see that the points lying on
generators of  $W_{2N - 1}(2)$ correspond to maximally commuting
subsets (MCSs) of operators and a spread of $W_{2N - 1}(2)$ is
nothing but a partition of the whole set of operators into
MCSs. Finally,
Eq.\,(9) tells us that there are $2^{2N-1}$ operators that do {\it
not} commute with a given operator.\footnote{Shortly after Ref. \cite{Saniga4} was posted on the arXiv-e, physicist D. Gross (Imperial College, London) sent us an outline of the proof of this property and a couple of weeks later, Koen Thas (Ghent University), a young mathematician, also informed us about finding a proof of the same statement.}

Recognizing $W_{2N - 1}(2)$ as the geometry behind $N$-qubits, we will now turn our attention on the properties
of the associated Pauli graphs,  $\mathcal{P}[2,N]$.

\subsection{Strong regularity of the $N$-qubit Pauli graph}
\label{partial}
\noindent
As already introduced in Sec.\,2.1, a strongly regular graph, srg$(v,D,\lambda,\mu)$, is a regular graph  having $v$ vertices and degree $D$ such that any two adjacent vertices are both adjacent to a constant number  $\lambda$ of vertices, and any two distinct non-adjacent vertices are also both adjacent to a constant number $\mu$ of vertices.  It is known that the adjacency matrix $A$ of any such graph satisfies the following equations \cite{DeClercq}
\begin{equation}
A J =D J,~~~~A^2+(\mu-\lambda)A+(\mu-D)I=\mu J,
\end{equation}
where $J$ is the all-one matrix. Hence, $A$ has $D$ as an eigenvalue with multiplicity one and its other eigenvalues are $r$ ($>0$) and $l$ ($<0$), related to each other as follows: $r+l=\lambda - \mu$ and $rl=\mu - D$. Strongly regular graphs exhibit many interesting properties \cite{DeClercq}. In particular, the two eigenvalues $r$ and $l$ are, except for (so-called) conference graphs, both integers, with
the following multiplicities 
\begin{equation}
f=\frac{-D(l+1)(D-l)}{(D+rl)(r-l)}~~\mbox{and}~~g=\frac{D(r+1)(D-r)}{(D+rl)(r-l)},
\end{equation}
respectively.
The $N$-qubit Pauli graph is strongly regular, and its properties can be inferred from the relation between symplectic polar spaces and partial geometries.

A partial geometry is a more general object than a finite generalized quadrangle.  It is finite near-linear space $\{P,L\}$ such that for any point $P$ not on a line $L$, (i) the number of points of $L$ joined to $P$ by a line equals $\alpha$, (ii) each line has $(s+1)$ points, (iii) each point is on $(t+1)$ lines; this partial geometry is usually denoted as pg$(s,t,\alpha)$ \cite{Batten}.
The graph of pg$(s,t,\alpha)$ is endowed with $v=(s+1)\frac{(st+\alpha)}{\alpha}$ vertices, $\mathcal{L}=(t+1)\frac{(st+\alpha)}{\alpha}$ lines and is strongly regular of the type
\begin{equation}
{\rm srg}\left((s+1)\frac{(st+\alpha)}{\alpha},s(t+1),s-1+t(\alpha -1),\alpha(t+1)\right).
\end{equation}
The other way round, if a strongly regular graph exhibits the spectrum of a partial geometry, such a graph is called a pseudo-geometric graph.  Graphs associated with symplectic polar spaces $W_{2N-1}(q)$ are pseudo-geometric \cite{DeClercq}, being
\begin{equation}
{\rm pg}\left(q\frac{q^{N-1}-1}{q-1},q^{N-1},\frac{q^{N-1}-1}{q-1}\right)\mbox{-graphs}.
\end{equation}
Combining these facts with the findings of the preceding section, we conclude that that $N$-qubit Pauli graph is of the type given by Eq.\,(13) for $q=2$; its basics invariants for a few small values of $N$ are listed in Table\,7. 
\begin{table}[h]
\begin{center}
\begin{tabular}{|r||r|r|r||r|r|r|r||r|r|r||}
\hline
$N$& $v$ & $\mathcal{L}$ & $D$ & $r$ & $l$ & $\lambda$ & $\mu$ & $s$ & $t$ & $\alpha$\\
\hline
\hline
$2$& $15$ & $15$ & $6$ & $1$ & $-3$ & $1$ & $3$ & $2$ & $2$ & $1$ \\
\hline
$3$& $63$ & $45$ & $30$ & $3$ & $-5$ & $13$ & $15$ & $6$ & $4$ & $3$\\
\hline
$4$& $255$ & $153$ & $126$ & $7$ & $-9$ & $61$ & $63$ & $14$ & $8$ & $7$\\
\hline
\end{tabular}
\label{simpleP22}
\caption{Invariants of the Pauli graph $\mathcal{P}[2,N]$, $N=2$, $3$ and $4$, as inferred from the properties of the symplectic polar spaces of order two and rank $N$. In general, $v=4^N -1$, $D=v-1-2^{2N-1}$, $s=2\frac{2^{N-1}-1}{2-1}$, $t=2^{N-1}$, $\alpha=\frac{2^{N-1}-1}{2-1}$, $\mu=\alpha(t+1)=rl+D$ and $\lambda=s-1+t(\alpha -1))=\mu+r+l$. The integers $v$ and $e$ can also be found from $s$, $t$ and $\alpha$ themselves.}
\end{center}
\end{table}

\section{The Pauli graph of two-qutrits}
\label{twoqutrits}
\noindent
A complete orthonormal set of operators of a single-qutrit Hilbert space is \cite{Lawrence2}
\begin{equation}
\sigma_I=\{ I_3,Z,X,Y,V,Z^2,X^2,Y^2,V^2\},~~I=1,2,3,\dots,9,
\end{equation}
where $I_3$ is the $3 \times 3$ unit matrix, $Z=\left(\begin{array}{ccc}1 & 0&0 \\0 & \omega&0\\0&0& \omega^2\\\end{array}\right)$, $X=\left(\begin{array}{ccc}0 & 0&1 \\1 & 0&0\\0&1& 0\\\end{array}\right)$, $Y=XZ$, $V=X Z^2$ and $\omega=\exp\left(2 i \pi/3\right)$.
Labelling the two-qutrit Pauli operators as follows $1=I_3 \otimes \sigma_1$, $2=I_3 \otimes \sigma_2$, $\cdots$, $8=I_3 \otimes \sigma_8$, $a=\sigma_1 \otimes I_3$, $9=\sigma_1 \otimes \sigma_1$,\ldots, $b=\sigma_2 \otimes I_3$, $17=\sigma_2 \otimes \sigma_1$,\ldots , $c=\sigma_3 \otimes I_3$,$\ldots$, $h=\sigma_8 \otimes I_2$,$\ldots$, $72=\sigma_8 \otimes \sigma_8$, one obtains the incidence matrix of the two-qutrit Pauli graph $\mathcal{P}_9$.

\begin{figure}[t]
\centerline{\includegraphics[width=7truecm,clip=]{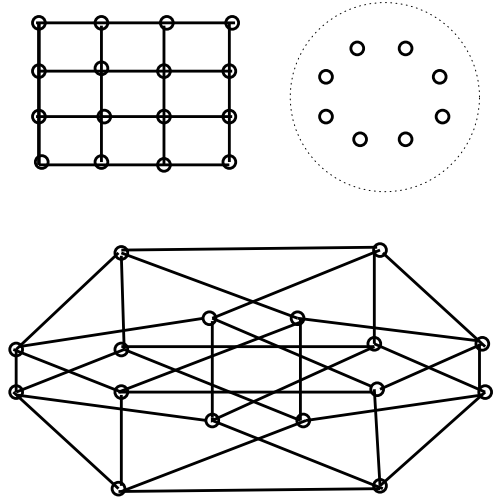}}
\caption{A partitioning of $\mathcal{W}_9$ into a grid (top left), an $8$-coclique
(top right) and a four-dimensional hypercube (bottom).}
\end{figure}
\begin{figure}[t]
\centerline{\includegraphics[width=9truecm,clip=]{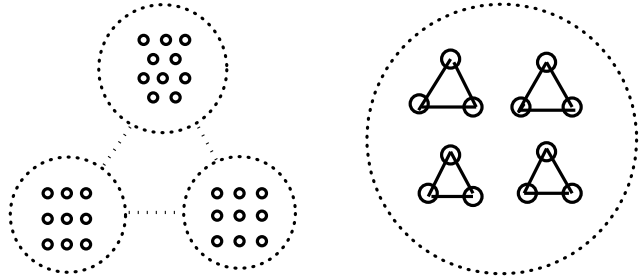}}
\caption{A partitioning of $\mathcal{W}_9$ into a tripartite graph comprising a $10$-coclique,
two $9$-cocliques and a set of four triangles; the lines corresponding to the vertices of a
selected triangle intersect at the same observables of $\mathcal{P}_9$ and the union of the
latter form a line of $\mathcal{P}_9$.}
\end{figure}
\begin{figure}[t]
\centerline{\includegraphics[width=8truecm,clip=]{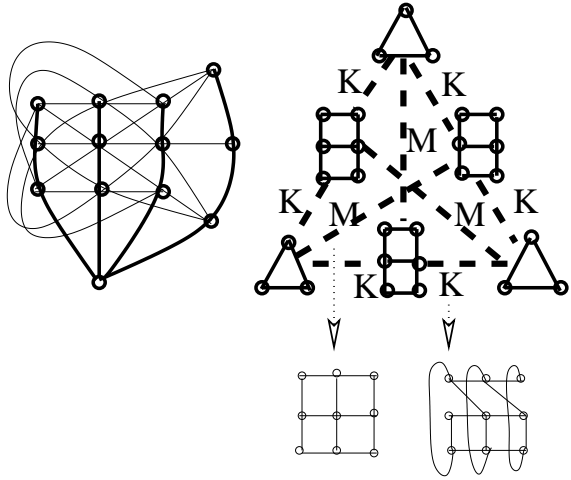}}
\caption{A partitioning of $\mathcal{W}_9$ into a perp-set and a ``single-vertex-sharing"
union of its three ovoids.}
\end{figure}

Computing the spectrum $\{-7^{15},-1^{40},5^{24},25\}$ one observes that the graph is regular, of degree $25$, but not strongly regular. The structure of observables in $\mathcal{P}_9$ is much more involved than in the case of two-qubits although it is still possible to recognize identifiable regular subgraphs.
In order to get necessary hints for the geometry behind this system, it necessitates
to pass to its dual graph, $\mathcal{W}_9$, i.\,e., the graph whose vertices are maximally commuting subsets
(MCSs) of  $\mathcal{P}_9$. To this end, let us first give a complete list of the latter:

\footnotesize
\begin{eqnarray}
&L_1=\{1,5,a,9,13,e,41,45\},~~L_2=\{2,6,a,10,14,e,42,46\},~~L_3=\{3,7,a,11,15,e,43,47\},\nonumber \\
&L_4=\{4,8,a,12,16,e,44,48\},~~M_1=\{1,5,b,17,21,f,49,53\},~~M_2=\{2,6,b,18,22,f,50,54\},\nonumber\\
&M_3=\{3,7,b,19,23,f,51,55\},~~M_4=\{4,8,b,20,24,f,52,56\},N_1=\{1,5,c,25,29,g,57,61\},\nonumber\\
&N_2=\{2,6,c,26,30,g,58,62\},~~N_3=\{3,7,c,27,31,g,59,63\},~~N_4=\{4,8,c,28,32,g,60,64\},\nonumber \\
&P_1=\{1,5,d,33,37,h,65,69\},~~P_2=\{2,6,d,34,38,h,66,70\},~~P_3=\{3,7,d,35,39,h,67,71\},\nonumber\\
&P_4=\{4,8,d,36,40,h,68,72\},\nonumber\\
&X_1=\{9,22,32,39,45,50,60,67\},~~X_2=\{10,17,27,40,46,53,63,68\},~~X_3=\{11,20,30,33,47,56,58,69\},\nonumber\\
&X_4=\{12,23,25,34,48,51,61,70\},X_5=\{13,18,28,35,41,54,64,71\},X_6=\{14,21,31,36,42,49,59,72\},\nonumber\\
&X_7=\{15,24,26,37,43,52,62,65\},X_8=\{16,19,29,38,44,55,57,66\},\nonumber\\
&Y_1=\{9,23,30,40,45,51,58,68\},~~Y_2=\{10,19,32,33,46,55,60,69\},~~Y_3=\{11,22,25,36,47,50,61,72\},\nonumber\\
&Y_4=\{12,17,26,39,48,53,62,67\},Y_5=\{13,20,27,34,41,56,63,70\},Y_6=\{14,23,28,37,42,51,64,65\},\nonumber\\
&Y_7=\{15,18,29,40,43,54,57,68\},Y_8=\{16,21,30,35,44,49,58,71\},\nonumber\\
&Z_1=\{9,24,31,38,45,52,59,66\},~~Z_2=\{10,24,25,35,46,52,61,71\},~~Z_3=\{11,17,28,38,47,53,64,66\},\nonumber\\
&Z_4=\{12,18,31,33,48,54,59,69\},Z_5=\{13,19,26,36,41,55,62,72\},Z_6=\{14,20,29,39,42,56,57,67\},\nonumber\\
&Z_7=\{15,21,32,34,43,49,60,70\},Z_8=\{16,22,27,37,44,50,63,65\}.\nonumber
\end{eqnarray}
\normalsize From there we find that $\mathcal{W}_9$ consists of 40
vertices and has spectrum $\{-4^{15},2^{24},12\}$, which are the
characteristics identical with those of the generalized quadrangle
of order three formed by the totally singular points and lines of
a parabolic quadric $Q(4,3)$ in $PG(4,3)$\cite{Payne}. The
quadrangle $Q(4,3)$, like its two-qubit counterpart, exhibits all
the three kinds of geometric hyperplanes, viz. a slim generalized
quadrangle of order (3,1) (a grid), an ovoid, and a perp-set, and
these three kinds of subsets can all indeed be found to sit inside
$\mathcal{W}_9$. One of the grids is formed by  the sixteen lines
$L_i$, $M_i$, $N_i$ and $P_i$ ($i=1$, $2$, $3$ and $4$) as
illustrated in Fig.\,9; the remaining 24 vertices comprise an
8-coclique ($X_i$, which correspond to mutually unbiased bases),
and a four-dimensional hypercube ($Y_i$ and $Z_i$). Next, one can
partition $\mathcal{W}_9$ into a maximum independent set and the
minimum vertex cover using a standard graph software. The
cardinality of any maximum independent set is 10 (= $3^2 + 1$), which
means that any such set is an ovoid of $Q(4,3)$\cite{Payne}. It is
easy to verify that, for example, the set $\{L_1,M_2,
N_3,P_4,X_3,X_8,Y_4,Y_6,Z_2,Z_7$\} is an ovoid; given any
maximum independent set, $Q(4,3)$/$\mathcal{W}_9$ can be partitioned
as shown in Fig.\,10. The remaining type of a hyperplane of
$\mathcal{W}_9$ is a perp-set, i.e. the set of 12 vertices
adjacent to a given (``reference") vertex (Fig.\,11); the set of
the remaining 27 vertices can be shown to consist of three ovoids
which share (altogether and pairwise) just a single vertex --- the
reference vertex itself. This configuration bears number $99$ in a
list of graphs with few eigenvalues given in Ref.
\cite{Evandam} and can schematically be illustrated in form of
a \lq\lq triangle", with a triangular pattern at its nodes and a
$1 \times 2$ grid put on its edges; the union of $1 \times 2$ grid
and a triangle either forms a Mermin-square-type graph $M$, as
already encountered in the two-qubit case, or a quartic graph of
another type, denoted as $K$ (see Fig.\,11).

The foregoing observations and facts provide a reliable basis for us to surmise that the
geometry behind $\mathcal{W}_9$ is identical with that of $Q(4,3)$. If this is so, then
the symplectic generalized quadrangle of order three, $W(3)$, which is the dual of
$Q(4,3)$\cite{Payne}, must underlie the geometry of the Pauli graph $\mathcal{P}_9$.
However, the vertex-cardinality of $W(3)$ is 40 (the same as that of $Q(4,3)$), whilst
$\mathcal{P}_9$ features as many as 80 points/vertices. Hence,  if the geometries of
$W(3)$ and $\mathcal{P}_9$ are isomorphic, then there must exits a natural pairing between
the Pauli operators such that there exists a bijection between pairs of operators of
$\mathcal{P}_9$ and points of $W(3)$. This issue requires, obviously, a much more elaborate
analysis, to be the subject of a separate paper\cite{multiline}.

\section{Conclusion}
\noindent
The paper introduces an important concept of the Pauli graph for the generalized Pauli operators of finite-dimensional quantum systems and illustrates and discussed this concept in an exhaustive detail for $N$-qubit systems, $N \geq 2$. In doing so, the geometries underlying these systems, viz. the symplectic polar spaces of rank $N$ and order two, are invoked to reveal all the intricacies of the algebra of the operators and its
basic factorizations.  Although there exits a variety of other interesting geometry-oriented
approaches to model finite dimensional quantum systems (see, for example, \cite{Aravind}--\cite{Klappenecker}), ours seems to be novel in that it goes beyond classical projective geometry and Galois fields and is, in principle, applicable to any quantum system of finite dimension.

\section*{Acknowledgements}
This work was partially supported by the
Science and Technology Assistance Agency under the contract $\#$
APVT--51--012704, the VEGA project $\#$ 2/6070/26 (both from
Slovak Republic) and the trans-national ECO-NET project $\#$
12651NJ ``Geometries Over Finite Rings and the Properties of
Mutually Unbiased Bases" (France). The second author also thanks Prof. Hans
Havlicek (Vienna University of Technology) for a number of enlightening discussions concerning
the structure of projective ring lines and their representations.

\end{document}